\documentclass[conference]{IEEEtran}
\IEEEoverridecommandlockouts
\usepackage{cite}
\usepackage{amsmath,amssymb,amsfonts}
\usepackage{graphicx}
\usepackage{textcomp}
\usepackage{xcolor}
\usepackage{algorithm}
\usepackage{algpseudocode}
\usepackage{subfigure}
\usepackage{balance} 
\def\BibTeX{{\rm B\kern-.05em{\sc i\kern-.025em b}\kern-.08em
    T\kern-.1667em\lower.7ex\hbox{E}\kern-.125emX}}

\newtheorem{lemma}{Lemma}

\newtheorem{remark}{Remark}

\newtheorem{prop}{Proposition}
\allowdisplaybreaks
\usepackage{xspace}
\usepackage{bbm}
\usepackage{mathrsfs}
\newcommand{\Dc}{\mathcal{D}}

\newcommand{\Fc}{\mathcal{F}}

\newcommand{\Hc}{\mathcal{H}}

\newcommand{\Mc}{\mathcal{M}}

\newcommand{\Oc}{\mathcal{O}}

\newcommand{\Rc}{\mathcal{R}}

\newcommand{\Mv}{{\bf M}}
\newcommand{\Nv}{{\bf N}}

\newcommand{\Qv}{{\bf Q}}

\newcommand{\Wv}{{\bf W}}
\newcommand{\Sv}{{\bf S}}

\newcommand{\muv}{\boldsymbol \mu}

\DeclareMathOperator\E{\mathbb E}
\let\P\relax
\DeclareMathOperator\P{\mathbb P}
\newcommand{\Norm}{\mathcal{N}}

\def\textiid{i.i.d.\@\xspace}
\newcommand\iid{\ifmmode\text{ i.i.d. } \else \textiid \fi}

\newcommand{\ind}{\boldsymbol{1}}

\newcommand{\indep}{\perp \!\!\! \perp}
\begin{document}

\title{Communication-Efficient Distribution-Free Inference Over Networks
}
 \author{%
   \IEEEauthorblockN{Mehrdad Pournaderi and Yu Xiang}
   \IEEEauthorblockA{University of Utah\\
   Department of  Electrical and Computer Engineering\\
                     50 Central Campus Dr 2110, Salt Lake City, USA\\
                     Email: \{m.pournaderi, yu.xiang\}@utah.edu}
 }
% \and
% \and
% \and
% \and
% \and

\maketitle
\begin{abstract}
Consider a star network where each local node possesses a set of test statistics that exhibit a symmetric distribution around zero 
%conditional on all the other (local) statistics 
when their corresponding null hypothesis is true.
%, motivated by the recent knockoff filter framework.
This paper investigates statistical inference problems in networks concerning the aggregation of this general type of statistics and global error rate control under communication constraints in various scenarios.
%depending on the structural relationships among the hypotheses, 
%under appropriate type-I error control constraints. 
The study proposes communication-efficient algorithms that 
%are statistically powerful against non-symmetric alternative distributions. These algorithms 
are built on established non-parametric methods, such as the Wilcoxon and sign tests, as well as modern inference methods such as the Benjamini-Hochberg (BH) and Barber-Cand{\`e}s (BC) procedures, coupled with sampling and quantization operations. The proposed methods are evaluated through extensive simulation studies.
\end{abstract}


\section{Introduction}
\noindent 
% Traditionally, the multiple testing procedures~\cite{benjamini1995,storey2002direct,efron2001empirical,genovese2002operating} are designed to achieve the false discovery rate (FDR) control based on the assumption that the p-values are available, for instance, the celebrated Benjamini-Hochberg (BH) procedure~\cite{benjamini1995}. In order to obtain the p-values, one needs to know the exact null distribution of each test statistic; this requirement, however, can be infeasible in practical settings. It is thus appealing to work with \emph{distribution-free} test statistics, where only mild structural assumptions are required. 

Statistical inference methods that do not assume a specific distribution or functional form for the data being analyzed are highly desirable in many practical applications. This has led to the development of non-parametric methods that rely on distribution-free or weakly-parametric assumptions, such as symmetry, independence, or exchangeability, to make inferences about the underlying distribution. Non-parametric methods are particularly useful when the underlying distribution of the data is unknown, difficult to model, or when the sample size is small. The broad applicability of non-parametric methods has made them popular in a variety of fields, including biology, economics, engineering, and social sciences~(see \cite{conover1999practical,hollander2013nonparametric} and references therein). 
%These methods offer several advantages over traditional parametric methods, including greater robustness to outliers, greater flexibility in modeling complex relationships, and the ability to handle data from a wide range of distributions without requiring extensive assumptions~\cite{hollander2013nonparametric}.}

 In many hypothesis testing scenarios, test statistics exhibit a symmetric distribution around zero under the null hypothesis. When multiple (independent) such test statistics are available, this symmetry property is sufficient for designing powerful non-parametric methods that are robust in nature without requiring extensive assumptions. Examples of such methods include 
%(non-parametric) sign test and 
Wilcoxon signed-rank test~\cite{bain1992introduction} for testing a single hypothesis and the Barber-Cand{\`e}s (BC) procedure~\cite{barber2015controlling} for multiple hypothesis testing. Importantly, these methods do not require the statistics to be independent, but only conditionally symmetric under the null hypothesis, i.e., the statistics (under the null) have a symmetric distribution around zero given all other statistics. For instance, the statistics computed using the recent knockoff filter framework~\cite{barber2015controlling} (also see various extensions~\cite{candes2018panning,barber2019knockoff,barber2020robust,lu2018deeppink,pournaderi2021differentially,pournaderi2022variable}) are not necessarily independent, but only conditionally symmetric (also referred to as the \iid sign property for the null  in~\cite{barber2015controlling}).

%This work concerns a class of \emph{distribution free} test statistics, motivated by the recent knockoff filter framework~\cite{barber2015controlling} (also see many extensions~\cite{candes2018panning,barber2019knockoff,barber2020robust,lu2018deeppink,pournaderi2021differentially,pournaderi2022variable} and references therein), where the signs of the null test statistics are conditionally symmetric (also referred to as the \iid sign property for the nulls in~\cite{barber2015controlling}). 

Inspired by various decentralized applications, we formulate the following distributed setting: Each node processes its own set of test statistics, and they wish to make decisions by communicating some information to a fusion center. We make an attempt to provide a comprehensive study by considering three different yet natural settings to cover a fairly general range of scenarios. The \emph{individual decision} setting, where each node wishes to make a decision on each of its own hypotheses (i.e., the multiple testing formulation) while achieving a global FDR control; the \emph{global decision} setting under the global null, for instance, all the nodes want to test for any signal given that they share the single null hypothesis; the \emph{distributed intersection hypothesis} setting that cover the cases where each node observes the test statistics for the same set of variables, and the goal is to collectively select the true variables under the FDR criterion. In the existing literature, perhaps the most related line of research concerns communication-efficient methods in distributed settings (e.g., wireless sensor networks) that assume the \emph{p-values} are available at each node~\cite{ermis2005adaptive,ray2007novel,ermis2009distributed,ray2011false,pournaderi2022sample,Ramdas2022,xiang2019distributed,pournaderi2022large}, while we relax this requirement by working directly with distribution-free test statistics.

Our proposed algorithms are built on distribution-free procedures such as the BC procedure~\cite{barber2015controlling}, as well as the sign tests and the Wilcoxon test to obtain p-values. For the global decision setting, we make use of the Simes' procedure~\cite{simes1986improved} for the global null, which is closely related to the BH procedure. We show that the communication complexity can be effectively accounted for by either quantization or sampling methods. We also carry out extensive numerical experiments to compare the algorithms with the pooled baselines (where all the test statistics are co-located). 

\section{Preliminaries}

\subsection{One-Sample Sign Test}
The one-sample sign test is a non-parametric hypothesis test that is used to test whether the median of a population is equal to a hypothesized value. To test $\mathsf{H}_0:\theta = 0$, one needs to count the number of observations with negative signs. Let $X$ denote this number and $n$ denote the total number of observations. Note that $X\sim \text{Bin}(n,1/2)$ under $\mathsf{H}_0$. Hence, the one-sided p-value to detect $\theta>0$ can be computed as $P =\P(\text{Bin}(n,1/2)\leq X) $.
\subsection{One-Sample Wilcoxon Signed-Rank Test\cite{bain1992introduction}}
The one-sample Wilcoxon signed-rank test is a non-parametric hypothesis test that is used to test whether the distribution of a population is symmetric around 0. For $n$ observations, let  $(S_1,\hdots,S_n)$ and $(R_1,\hdots,R_n)$ denote the signs and ranks of the absolute value of the observations, respectively. The Wilcoxon statistics is defined as $W = \sum_{i=1}^n R_i\ind\{S_i>0\}$. Under the null hypothesis (i.e., conditionally symmetric distribution for all observations), $\E W = \frac{1}{4} n (n+1)$, $\text{var}(W) = \frac{1}{24}n(n+1)(2n+1)$ and $\frac{W-\E W}{\sqrt{\text{var}(W)}}\xrightarrow{\mathcal{D}}\mathcal{N}(0,1)$ which allows us to compute asymptotic p-values.

\subsection{Barber-Cand{\`e}s (BC) Procedure \cite{barber2015controlling}}
\label{BC}
The BC procedure makes a decision for each observation $W_i$, $i\in [n]$ where $[n]=\{1,\hdots,n\}$, testing whether it is generated from a conditionally symmetric distribution. The procedure controls a type-I error rate called false discovery rate (FDR) at a prespecified level $\alpha$. Let $R$ and $V$ denote the number of rejections and false rejections, respectively. FDR is defined as the expected value of the false discovery proportion (FDP), i.e.,  $\mathsf{FDR}=\mathbb{E}(\mathsf{FDP})$ with $\mathsf{FDP} = \frac{V}{R\vee 1}$, ($a\vee b:=\max\{a,b\}$).
The BC procedure is a threshold method rejecting $\{j:W_j\geq T\}$, where 
\begin{equation*}
    T := \min\{t\in \Psi: \widehat{\mathsf{FDP}}(t) \leq \alpha\}
\end{equation*}
with $\Psi:=\{|W_i|:i\in[n]\}\backslash \{0\}$, $\widehat{\mathsf{FDP}}(t):= \frac{\widehat{V}(t)}{R(t)\vee 1}$, $\widehat{V}(t): = 1+\sum_{i=1}^n \ind\{W_i\leq -t\}$, and $R(t):=\sum_{i=1}^n \ind\{W_i\geq t\}$.

\subsection{Benjamini-Hochberg (BH) Procedure \cite{benjamini1995}}
For $n$ independent p-values, the BH procedure
makes a decision for each p-value $P_i$, $i\in [n]$, testing whether it is generated from a super-uniform distribution, i.e., $\P(P_i\leq t)\leq t$. The procedure rejects $R_{\text{BH}}:=\max\{k:P_{(k)}\leq k\alpha/n\}$ smallest p-values (with convention $\max \varnothing = 0$) where $P_{(k)}$ denotes the $k$-th smallest p-value. It controls the FDR at some prefixed level $\alpha$.

\section{Problem Settings}
Let $(\mathsf{H}_1^{(i)},\hdots,\mathsf{H}_{m^{(i)}}^{(i)})$ denote a tuple of hypotheses and $\Wv^{(i)}=(W_1^{(i)},\hdots,W_{m^{(i)}}^{(i)})$ a vector of the corresponding statistics at node $i$, $i\in [N]$, where $[N]=\{1,\hdots,N\}$. Let $-j$ denote the set of indices without $j$, namely, $-j:=[N]\setminus\{j\}$. We assume that statistics corresponding to different nodes are independent, i.e., $\Wv^{(i)}\indep \Wv^{(j)}$, for all $i\neq j$. At each node $i$, we assume %$\P(W_j^{(i)}\neq 0)>0$ and 
that the conditional distribution of $W_j^{(i)}$ given $W_{-j}^{(i)}$, i.e., $W_j^{(i)}\big|W_{-j}^{(i)}$ is symmetric for all $j\in\Hc_0^{(i)}$, where $\Hc_0^{(i)}=\{j:\mathsf{H}_j^{(i)}\ \text{is true}\}$, i.e., 
\begin{align}
    W_j^{(i)}\big|W_{-j}^{(i)}\overset{d}{=}-W_j^{(i)}\big|W_{-j}^{(i)},\label{eq:sym}
\end{align}
for $j\in\Hc_0^{(i)}$. Note that this is equivalent to the \iid sign property for the nulls in~\cite{barber2015controlling} (also see~\cite{barber2020robust}). In our distributed settings, we will refer to this property as the \emph{local} conditional symmetry. We consider the following settings. 
%the sign of the null statistics are independent of magnitudes $\Mv^{(i)}=\left(|W_1^{(i)}|,\hdots,|W_{m^{(i)}}^{(i)}|\right)$ and distributed as $\iid$ Rademacher random variables.

%Let $R$ and $V$ denote the number of rejections and false rejections by some procedure, respectively. FDR is defined as the expected value of the false discovery proportion (FDP), i.e.,  $\mathsf{FDR}=\mathbb{E}(\mathsf{FDP})$ with $\mathsf{FDP} = \frac{V}{R\vee 1}$, ($a\vee b:=\max\{a,b\}$).

\smallskip

{\noindent \underline{\textbf{Setting I: individual decision:}}}
In this setting, a separate decision is made for each hypothesis $\mathsf{H}_j^{(i)},\ i\in[N],\ j\in[m^{(i)}]$ in the network under the global FDR control constraint. Two methods are proposed, namely, \emph{pooled q-BC} and \emph{sampled BC} procedures which utilize quantization and sampling operations to attain $\Oc(m)$ and $\Oc(\log m)$ bits communication complexity, respectively.
\smallskip

{\noindent \underline{\textbf{Setting II: global decision:}}}
In this setting, the goal is to test the single (global) hypothesis $\bigcap_{i=1}^N\bigcap_{j=1}^{m^{(i)}}\mathsf{H}_j^{(i)}$ with a control over the probability of false rejection. We provide two methods for each of the three classes based on the communication complexity: $\Oc(m)$, $\Oc(\log m)$, and $\Oc(1)$.
\smallskip
 
% {\noindent \underline{\textbf{Local intersection hypothesis:}}}

% \smallskip

{\noindent \underline{\textbf{Setting III: distributed intersection hypothesis:}}}
In this setting, we consider $m^{(i)}=m'$ for all $i\in [N]$ and a separate decision is made for each $\mathsf{H}_j=\bigcap_{i=1}^N\mathsf{H}_j^{(i)}$ under the FDR control constraint. We propose a simple method with $\Oc(m)$ bits communication complexity which outperforms an existing method~\cite{su2015communication}.
\smallskip

The tests are presented in the one-sided form where they are powerful against large positive alternatives. However, the arguments can be extended to two-sided tests via Bonferroni correction. 
\section{Individual Decision Setting}
In this section, we propose two algorithms for the individual decision setting. Let 
\begin{align*}
\Sv^{(i)}&=\left(S_1^{(i)},\hdots,S_{m^{(i)}}^{(i)}\right):=\mathsf{sgn}\left(\Wv^{(i)}\right)\text{ and }\\
\Mv^{(i)}&=\left(|W_1^{(i)}|,\hdots,|W_{m^{(i)}}^{(i)}|\right)
\end{align*}
denote the vector of signs and magnitudes of the statistics at node $i$, respectively.

\subsection{Pooled q-BC algorithm:}
This algorithm relies on quantizing the normalized magnitudes $\frac{\Mv^{(i)}}{\|\Mv^{(i)}\|_\infty}$ of the statistics for communication efficiency and has $\Oc(m)$ bits communication complexity.
\begin{algorithm}
\caption{Pooled q-BC}
    \textbf{Input:} statistics vectors $\Wv^{(1)},\hdots,\Wv^{(N)}$ and target FDR $\alpha$\\ 
     \textbf{Output:} rejected hypotheses at each node\\
    \textbf{each node $i$:}
    \begin{algorithmic}[1]
        \State\text{quantize:}
         $\Qv^{(i)}\gets\mathsf{quant}^{(i)}\left(\frac{\Mv^{(i)}}{\|\Mv^{(i)}\|_\infty}\right)$;
         \State\text{compute:}    $\Wv^{(i)}_{\text{q}}\gets\left(S_1^{(i)}Q_1^{(i)},\hdots,S_{m^{(i)}}^{(i)}Q_{m^{(i)}}^{(i)}\right)$;
         \State\text{transmit $\Wv^{(i)}_{\text{q}}$ to the center node};
    \end{algorithmic}
         \textbf{center node:} 
    \begin{algorithmic}[1]
         \State pool $\Wv^{(i)}_{\text{q}}$ from all the nodes and apply the BC procedure;
         \State broadcast the BC threshold $T_{\mathsf{qBC}}$;
    \end{algorithmic}
         \textbf{each node $i$:} 
    \begin{algorithmic}[1]      
         \State reject $\Rc^{(i)}=\{j:W_j^{(i)}\geq T_{\mathsf{qBC}}\}$;
    \end{algorithmic}
    \label{alg:pooled q-BC}
\end{algorithm}
%Let $f^{(i)}:\mathbb R_{\geq 0}^{m^{(i)}}\to\mathbb Q_{\geq 0}^{m^{(i)}}$ denote the general quantization operator at node $i$, i.e., $\Qv^{(i)}:=f^{(i)}\left({\Mv^{(i)}}\right)$. 
%$f^{(i)}(\cdot)$.
%We require $f^{(i)}$ to map the zero elements of $\Mv^{(i)}$ to zero, i.e., if $M^{(i)}_j=0$, then $Q^{(i)}_j=0$. 
% In other words, letting $\Qv^{(i)}:=f^{(i)}\left({\Mv^{(i)}}\right)$, the only necessary assumption on  is to be independent of $\Sv^{(i)}$.
It should be noted that we make no assumptions on
the quantization operator, i.e., $\Qv^{(i)}$ can be any function of $\Mv^{(i)}$ 
%(independent of $\Sv^{(i)}$) 
and it can vary throughout the network.
\smallskip
\begin{prop}
The pooled q-BC procedure controls the FDR globally. \label{thm:qbc}
\end{prop}
\smallskip
\begin{IEEEproof}
% {\color{blue} Since the nodes are assumed to be independent local conditional symmetry of the null statistics implies the global conditional symmetry. We note that the conditional symmetry condition is equivalent to having sign of the non-zero null statistics \iid Rademacher and independent of the set of magnitudes. According to the proof in \cite{barber2015controlling}, this is sufficient for FDR control. We must show that this property is preserved under quantization. But, since quantization maps zero statistics to zero and does not use the sign information we still have the sign of the non-zero null statistics \iid Rademacher and independent of the set of magnitudes. Therefore, the claim follows. }
According to the proof in \cite{barber2015controlling}, it is sufficient to prove the \emph{global} conditional symmetry, i.e.,
\begin{multline}
   \P\left(W_{q,j}^{(i)}>0\,\bigg | \left|W_{q,j}^{(i)}\right|,W_{q,-j}^{(i)},\Wv_{q}^{(-i)}\right) =\\
   \P\left(W_{q,j}^{(i)}<0\,\bigg | \left|W_{q,j}^{(i)}\right|,W_{q,-j}^{(i)},\Wv_{q}^{(-i)}\right)\ a.s.,\label{eq:sym}
\end{multline}
for all $i\in[N],\ j\in\Hc_0^{(i)}$, which reduces to  
\begin{multline*}
   \P\left(W_{q,j}^{(i)}>0\,\bigg | \left|W_{q,j}^{(i)}\right|,W_{q,-j}^{(i)}\right) =\\
   \P\left(W_{q,j}^{(i)}<0\,\bigg | \left|W_{q,j}^{(i)}\right|,W_{q,-j}^{(i)}\right)\ a.s.\, 
\end{multline*}
since the nodes are assumed to be independent. Let $\Sv_q^{(i)}=\mathsf{sgn}\big(\Wv_q^{(i)}\big)$. By the definitions,  $\big(|W_{q,j}^{(i)}|,W_{q,-j}^{(i)}\big)$ is a function of $(\Qv^{(i)},S_{q,-j}^{(i)})$. Hence, it is sufficient to prove
\[
    \P\left(S_{q,j}^{(i)}=1\,\bigg | \Qv^{(i)},S_{q,-j}^{(i)}\right)=\P\left(S_{q,j}^{(i)}=-1\,\bigg | \Qv^{(i)},S_{q,-j}^{(i)}\right)
\]
almost surely.
%According to the definition of $W_{q,j}^{(i)}$, we observe that the two $\sigma$-algebras   $\sigma\left(|W_{q,j}^{(i)}|,W_{q,-j}^{(i)}\right)$ and  $\sigma(\Qv^{(i)},S_{q,-j}^{(i)})$ are equivalent, which implies that 
% \begin{align*}
% \P \left(W_{q,j}^{(i)}>0\,\bigg | \left|W_{q,j}^{(i)}\right|,W_{q,-j}^{(i)}\right)=
%     \P\left(S_{q,j}^{(i)}=1\,\bigg | \Qv^{(i)},S_{q,-j}^{(i)}\right).
% \end{align*}
Now we note 
\begin{align*}
    &\P\left(S_{q,j}^{(i)}=1\,\bigg | \Qv^{(i)},S_{q,-j}^{(i)}\right)\\
     &=\P\left(S_{j}^{(i)}=1,Q_j^{(i)}\neq 0\,\bigg | \Qv^{(i)},S_{q,-j}^{(i)}\right)\\
    &=\ind\{Q_j^{(i)}\neq 0\}\P\left(S_{j}^{(i)}=1\,\bigg | \Qv^{(i)},S_{q,-j}^{(i)}\right)\\
    &=\ind\{Q_j^{(i)}\neq 0\}\E\left\{\P\left(S_{j}^{(i)}=1\,\bigg | \Mv^{(i)},S^{(i)}_{-j}\right)\bigg|\Qv^{(i)},S_{q,-j}^{(i)}\right\}\\
    &\overset{(*)}{=}\ind\{Q_j^{(i)}\neq 0\}\E\left\{\P\left(S_{j}^{(i)}=-1\,\bigg | \Mv^{(i)},S^{(i)}_{-j}\right)\bigg|\Qv^{(i)},S_{q,-j}^{(i)}\right\}\\
    &=\ind\{Q_j^{(i)}\neq 0\}\P\left(S_{j}^{(i)}=-1\,\bigg | \Qv^{(i)},S_{q,-j}^{(i)}\right)\\
    %&=\P\left(S_{j}^{(i)}=-1,Q_j^{(i)}\neq 0\,\bigg | \Qv^{(i)},S_{q,-j}^{(i)}\right)\\
    &=
    \P\left(S_{q,j}^{(i)}=-1\,\bigg | \Qv^{(i)},S_{q,-j}^{(i)}\right)\quad a.s.,
    %\\
     %&=\P \left(W_{q,j}^{(i)}<0\,\bigg | \left|W_{q,j}^{(i)}\right|,W_{q,-j}^{(i)}\right)\quad a.s.,
\end{align*}
where $(*)$ holds according to the conditional symmetry of the original null statistics, completing the proof.
\end{IEEEproof}

\subsection{Sampled BC algorithm:}
This algorithm relies on sampling the $\widehat V(t)$ and $R(t)$ processes (defined in Section \ref{BC}) for communication efficiency and has $\Oc(\log m)$ bits communication complexity.
\begin{algorithm}
\caption{Sampled BC}
    \textbf{Input:} statistics vectors $\Wv^{(1)},\hdots,\Wv^{(N)}$ and target FDR $\alpha$,\\ 
    \text{\qquad\ \ } number of samples at each node: $L\geq 2$,\\
             \text{\qquad\ \ } sampling locations: $t_\ell=\frac{\ell-1}{L-1},\ \ell\in [L]$.\\
    \textbf{Output:} rejected hypotheses at each node\\
    \textbf{each node $i$:}
    \begin{algorithmic}[1]
        \State\text{normalize:}
         $\Nv^{(i)}\gets\frac{\Wv^{(i)}}{\|\Mv^{(i)}\|_\infty}$;
         \For{$\ell\in [L]$}
             \State\text{sample:}    $\widehat V_\ell^{(i)} \gets \widehat V^{(i)}(t_\ell)=\sum_{j=1}^{m^{(i)}}\ind\{N_j^{(i)} < -t_\ell\}$;
              \State \qquad\quad\    $R_\ell^{(i)} \gets R^{(i)}(t_\ell)=\sum_{j=1}^{m^{(i)}}\ind\{N_j^{(i)} > t_\ell\}$;
         \EndFor
         \State\text{transmit $(\widehat V_\ell^{(i)},R_\ell^{(i)})_{\ell\in [L]}$ to the center node};
    \end{algorithmic}
         \textbf{center node:} 
    \begin{algorithmic}[1]
         \State pool: $\widehat{\mathsf{FDP}}_\ell\gets\frac{1+\sum_{i=1}^N \widehat{V}_\ell^{(i)}}{1\vee \sum_{i=1}^N R_\ell^{(i)}},\ \ell\in [L]$ ;
         \State compute and broadcast the BC threshold index $K_{\mathsf{sBC}}= \min\left\{\ell\in [L]:\,{\widehat{\mathsf{FDP}}}_\ell\leq \alpha\right\}$, $\min(\varnothing):=L$;
    \end{algorithmic}
         \textbf{each node $i$:} 
    \begin{algorithmic}[1]      
         \State reject $\Rc^{(i)}=\{j:N_j^{(i)}> T_{\mathsf{sBC}}:=t_{K_{\mathsf{sBC}}}\}$;
    \end{algorithmic}
    \label{alg:samp BC}
\end{algorithm}

\smallskip
\begin{prop}
The sampled BC procedure controls the FDR globally.
\end{prop}
\smallskip
\begin{IEEEproof}
According to the proof of the original BC procedure in \cite{barber2015controlling}, we have
\[
    \text{FDR}\leq \alpha \cdot
    \E\left(\frac{\sum_{i=1}^N V_{+,K_{\mathsf{sBC}}}^{(i)}}{1+\sum_{i=1}^N V_{-,K_{\mathsf{sBC}}}^{(i)}}\right).
\]
where $V_{+,K_{\mathsf{sBC}}}^{(i)}:=\sum_{j\in\Hc_0^{(i)}}\ind\{N_j^{(i)}> T_{\mathsf{sBC}}\}$ and $V_{-,K_{\mathsf{sBC}}}^{(i)}:=\sum_{j\in\Hc_0^{(i)}}\ind\{N_j^{(i)}< -T_{\mathsf{sBC}}\}$.
Hence, it is sufficient to prove
\begin{equation*}
    \E\left(\frac{\sum_{i=1}^N V_{+,K_{\mathsf{sBC}}}^{(i)}}{1+\sum_{i=1}^N V_{-,K_{\mathsf{sBC}}}^{(i)}}\right)\leq 1.
\end{equation*}
We note,
\begin{equation*}
    \frac{\sum_{i=1}^N V_{+,\ell}^{(i)}}{1+\sum_{i=1}^N V_{-,\ell}^{(i)}}
    %=\frac{V_{+,\ell}}{1+V_{-,\ell}}
    =\frac{V_+(t_\ell)}{1+V_-(t_\ell)},
\end{equation*}
where $V_+(t_\ell)=\sum_{i=1}^N \sum_{j\in\Hc_0^{(i)}}\ind\{N_j^{(i)}> t_\ell\}$ and $V_-(t_\ell)=\sum_{i=1}^N \sum_{j\in\Hc_0^{(i)}}\ind\{N_j^{(i)}< -t_\ell\}$.
Using a similar argument as in the proof of the BC procedure, we observe that $\Mc(t)=\frac{V_+(t)}{1+V_-(t)},\ t\geq 0$ forms a supermartingale w.r.t. the filtration 
\begin{align*}
    \Fc(t)=\sigma\Big(\big\{(V_+(s),V_-(s))&:0\leq s\leq t\big\}, \big\{N_j^{(i)}:\mathsf{H}_j^{(i)}\ \text{false}\big\},\\
    & \big\{\big|N_j^{(i)}\big|:i\in[N],\ j\in[m^{(i)}]\big\}\Big)
\end{align*}
%$\Fc(t)=\sigma(V_+(s),V_-(s),\ind\{\mathsf{H}_j^{(i)}\ \text{is true},|N_j^{(i)}|\leq s\};0\leq s\leq t)$ 
and $\E\left(\frac{V_+(0)}{1+V_-(0)}\right)\leq 1$. 
%Since $t_\ell\leq 1$ for all $\ell\in [L]$, the optional sampling theorem\footnote{{\color{red}Although $\Mc(t)$ is c\' adl\' ag by construction, we do not need it to invoke the optional sampling theorem since }} implies that 
Therefore, $(\Mc(t_\ell),\Fc(t_\ell))_{\ell\in[L]}$ is also a supermartingale and by the optional stopping theorem, we get $\E\left(\frac{V_{+}(T_{\mathsf{sBC}})}{1+V_{-}(T_{\mathsf{sBC}})}\right)\leq 1$, completing the proof.  
\end{IEEEproof}
\section{Global Decision Setting}
In this section, we wish to present algorithms for testing the global hypothesis $\bigcap_{i=1}^N\bigcap_{j=1}^{m^{(i)}}\mathsf{H}_j^{(i)}$. We will use the following lemma from \cite{benjamini1995}.
\begin{lemma}[\!\!\cite{benjamini1995}]
Let $R$ and $V$ denote the number of rejections and false rejections by some $\alpha$-level FDR controlling procedure applied to $\mathsf{H}_j^{(i)},\ i\in[N],\ j\in[m^{(i)}]$. If $\bigcap_{i=1}^N\bigcap_{j=1}^{m^{(i)}}\mathsf{H}_j^{(i)}$ is true, then $\P(R> 0)\leq \alpha$.
\label{lem:glob}
\end{lemma}
\begin{IEEEproof}
Observe that 
\begin{equation*}
    \mathsf{FDR} = \E\left(\frac{V}{R\vee 1}\Big|R> 0\right)\P(R> 0)=\P(R> 0)
\end{equation*}
%$\E\left(\frac{V}{R\vee 1}\Big|R> 0\right)\P(R> 0)=\P(R> 0)$ 
since $V=R$ under the global null. 
Therefore, $\P(R> 0)\leq \alpha$ according to $\mathsf{FDR}\leq\alpha$ . 
%We have $\alpha \geq \mathsf{FDR} =\P(R> 0)=\P(V> 0)$. 
\end{IEEEproof}

This also follows from the well-known fact that FDR is equivalent to FWER (i.e., $P(V>0)$) under the global null. An immediate implication of this lemma is that given an FDR controlling procedure for $\mathsf{H}_j^{(i)},\ i\in[N],\ j\in[m^{(i)}]$, one can test the global null hypothesis $\mathsf{H}_g=\bigcap_{i=1}^N\bigcap_{j=1}^{m^{(i)}}\mathsf{H}_j^{(i)}$ by rejecting $\mathsf{H}_g$ if $R>0$. The Simes p-value \cite{simes1986improved} is an example of this approach\cite{10.2307/26773160}.
\subsection{$\Oc(m)$ methods:}
\subsubsection{pooled q-BC} This method is simply applying Algorithm~\ref{alg:pooled q-BC} without the last two step. Instead, we reject $\mathsf{H}_g$ if any hypothesis is rejected, i.e., $R>0$ or equivalently if $T_{\mathsf{qBC}}<\infty$. This rejection rule is statistically valid according to Lemma~\ref{lem:glob}.
\subsubsection{Wilcoxon signed-rank test} This method also applies Algorithm~\ref{alg:pooled q-BC} except for the last two steps. According to the proof of Proposition~\ref{thm:qbc}, pooled statistics satisfy the global conditional symmetry property. Hence the Wilcoxon signed-rank test can be applied to test the global null and it is asymptotically valid.
\subsection{$\Oc(\log m)$ methods:}
\subsubsection{Sign test}
In this method each node sends the number of statistics and negative statistics to the center node. The center node computes the total number of statistics and negative statistics and then applies the one-sample sign test to decide about the global hypothesis $\bigcap_{i=1}^N\bigcap_{j=1}^{m^{(i)}}\mathsf{H}_j^{(i)}$.
\subsubsection{Sampled BC}
This method is simply applying Algorithm~\ref{alg:samp BC} without the last two step. Instead, the global null is rejected if $\underset{\ell\in [L]}{\min}\,\widehat{\mathsf{FDP}}_\ell\leq \alpha$. This rejection rule is statistically valid according to Lemma~\ref{lem:glob}.
\subsection{$\Oc(1)$ methods:}
\subsubsection{Wilcoxon+Simes (Algorithm~\ref{alg:W+S})}
In this method a local p-value is computed by applying the Wilcoxon test at each node. The p-values then are quantized and transmitted to the center node where they are fused and a final decision is made. Notice that the method is only asymptotically valid since the Wilcoxon p-values are computed according to the asymptotic distribution of the statistic under the global null.
\begin{algorithm}
\caption{Wilcoxon+Simes}
    \textbf{Input:} statistics vectors $\Wv^{(1)},\hdots,\Wv^{(N)}$, target FDR $\alpha$, the 
    number of quantization levels for each node $k^{(1)},\hdots,k^{(N)}$
     \textbf{Output:} rejected hypotheses at each node\\
    \textbf{each node $i$:}
    \begin{algorithmic}[1]
        \State\text{Wilcoxon:} compute a local p-value $P^{(i)}$ by applying the Wilcoxon test;
        \State\text{quantize:}
         $Q^{(i)}\gets\frac{1}{ k^{(i)}}\lceil k^{(i)} P^{(i)}\rceil$;
         \State\text{transmit $Q^{(i)}$ to the center node};
    \end{algorithmic}
         \textbf{center node:} 
    \begin{algorithmic}[1]
         \State compute the Simes p-value $S\gets\underset{i\in [N]}{\min}\frac{Q_{(i)}\cdot N}{i}$ where $Q_{(i)}$ is the $i$-th order statistics for $(Q^{(i)})_{i\in [N]}$;
         \State reject the global hypothesis if $S\leq \alpha$;
    \end{algorithmic}
    \label{alg:W+S}
\end{algorithm}

% \smallskip

% \begin{fact}
% If $P^{(i)}$ is super-uniform, i.e., $\P(P^{(i)}\leq t)\leq t$ then $Q^{(i)}=\frac{1}{ k^{(i)}}\lceil k^{(i)} P^{(i)}\rceil$ is superuniform.
% \end{fact}
% \smallskip

% %\begin{IEEEproof}
% This follows since 
% \begin{align*}
%         \P\left(\frac{1}{ k^{(i)}}\lceil k^{(i)} P^{(i)}\rceil\leq t\right)&=\P\left( k^{(i)} P^{(i)}\leq \lfloor k^{(i)} t \rfloor\right)\\
%         &\leq \P(P^{(i)}\leq t)\leq t.
%     \end{align*}
% %\end{IEEEproof}
The following proposition concerns the asymptotic validity of the quantized p-values computed in Algorithm~\ref{alg:W+S}.
\begin{prop}
If $P_{m^{(i)}}\xrightarrow{\Dc}SU[0,1]$ (where $SU$ stands for a superuniform distribution), i.e., $\lim_{m^{(i)}\to\infty}\P(P_{m^{(i)}}\leq t)\leq t$, $t\geq 0$, then $Q_{m^{(i)}}=\frac{1}{ k^{(i)}}\lceil k^{(i)} P_{m^{(i)}}\rceil\xrightarrow{\Dc}SU[0,1],\ k^{(i)}>0$.
\end{prop}
\begin{IEEEproof}
We note
\begin{align*}
    \P&\left(Q_{m^{(i)}}\leq t\right)=\P\left(\frac{1}{ k^{(i)}}\lceil k^{(i)} P_{m^{(i)}}\rceil\leq t\right)\\
    &=\P\left( k^{(i)} P_{m^{(i)}}\leq \lfloor k^{(i)} t \rfloor\right)=\P\left( P_{m^{(i)}}\leq \lfloor k^{(i)} t \rfloor/k^{(i)}\right).
\end{align*}
Therefore, 
\begin{align*}
\lim_{m^{(i)}\to\infty}\P\left(Q_{m^{(i)}}\leq t\right) &= \lim_{m^{(i)}\to\infty}\P\left( P_{m^{(i)}}\leq \lfloor k^{(i)} t \rfloor/k^{(i)}\right)\\
&\leq \lfloor k^{(i)} t \rfloor/k^{(i)} \leq t,
\end{align*}
concluding the claim.
\end{IEEEproof}
\smallskip

\begin{remark}
The communication cost of Algorithm~\ref{alg:W+S} can be further reduced by transmitting only $\{Q^{(i)}\leq \alpha\}$ ($\min \varnothing = 1$).
\end{remark}

\subsubsection{Sign test+Simes}
This method is the same as Algorithm~\ref{alg:W+S}, except it uses the one-sample sign test to compute the local p-values instead of Wilcoxon test. This algorithm is statistically valid in finite-sample sense since the p-values computed according to the sign statistics are exact.
% \section{Local intersection hypothesis}

\begin{figure*}[t]\centering
\begin{subfigure}{}\includegraphics[scale=0.4]{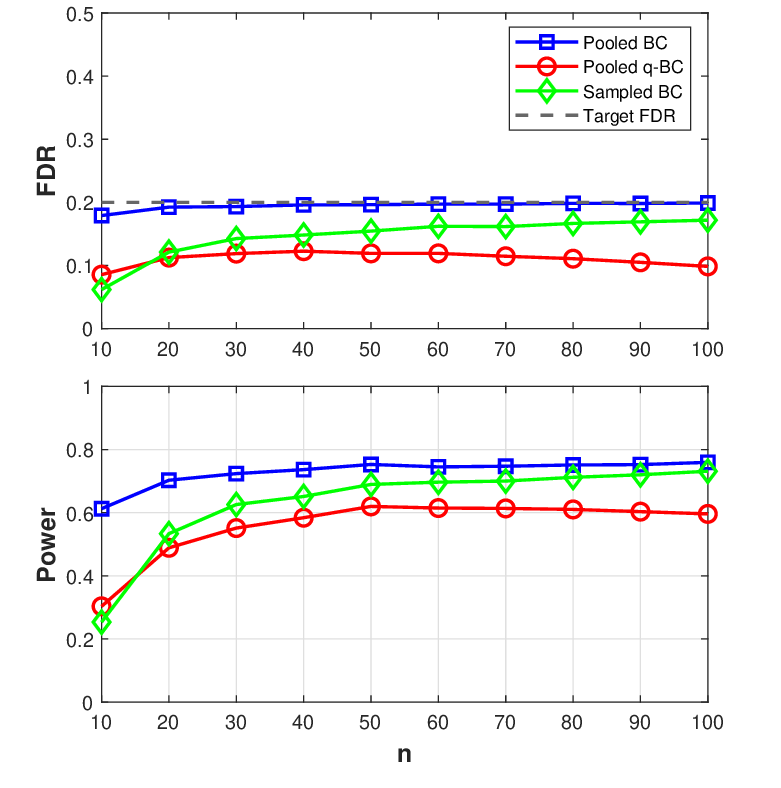}\end{subfigure}
\begin{subfigure}{}\includegraphics[scale=0.4]{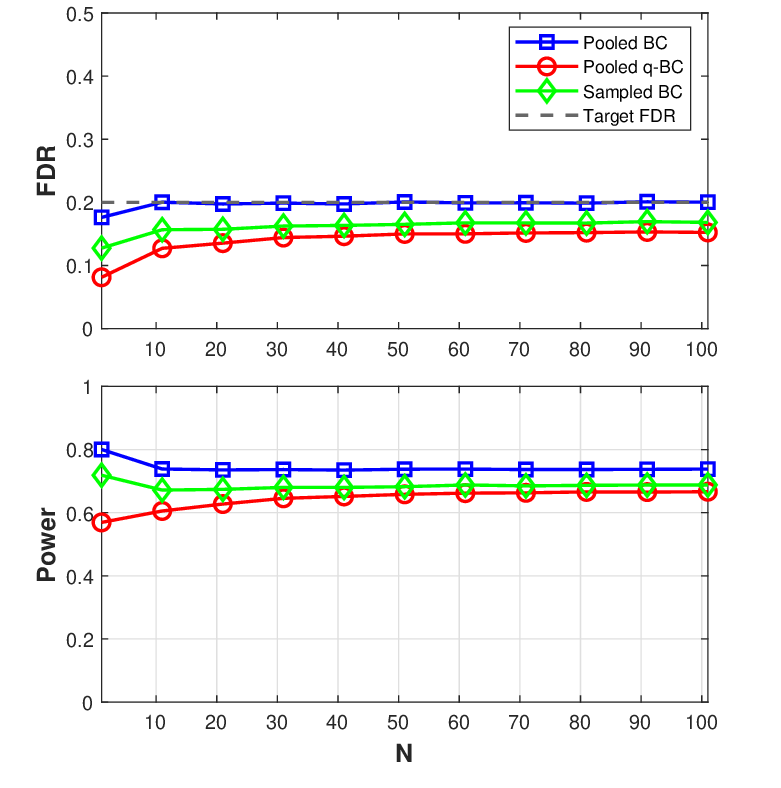}\end{subfigure}
\begin{subfigure}{}\includegraphics[scale=0.4]{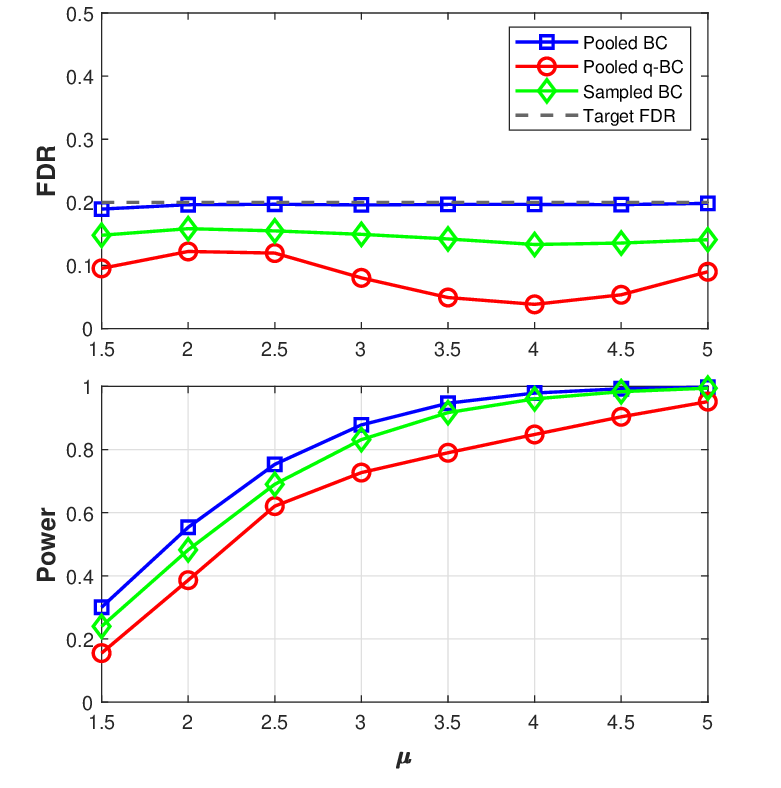}\end{subfigure}
\caption{Experiment~1. From left to right, Simulations I, II, and III.}
\vspace{-1em}
\label{fig:individual}
\end{figure*}

\section{Distributed Intersection Hypothesis Setting}
In this section we present an algorithm of $\Oc(m)$ communication complexity for testing $\mathsf{H}_j=\bigcap_{i=1}^N\mathsf{H}_j^{(i)}$. 
%Let $\overline{\Wv}=\frac{1}{N}\sum_{i=1}^N \Wv^{(i)}$.
\begin{algorithm}
\caption{Averaged BC}
    \textbf{Input:} statistics vectors $\Wv^{(1)},\hdots,\Wv^{(N)}$ and target FDR $\alpha$\\ 
     \textbf{Output:} rejected hypotheses at each node\\
    \textbf{each node $i$:}
    \begin{algorithmic}[1]
        \State\text{quantize:}
         $\Qv^{(i)}\gets\mathsf{quant}^{(i)}\left(\frac{\Mv^{(i)}}{\|\Mv^{(i)}\|_\infty}\right)$;
         \State\text{compute:}    $\Wv^{(i)}_{\text{q}}\gets\left(S_1^{(i)}Q_1^{(i)},\hdots,S_{m^{(i)}}^{(i)}Q_{m^{(i)}}^{(i)}\right)$;
         \State\text{transmit $\Wv^{(i)}_{\text{q}}$ to the center node};
    \end{algorithmic}
         \textbf{center node:} 
    \begin{algorithmic}[1]
         \State compute $\overline{\Wv}=\frac{1}{N}\sum_{i=1}^N \Wv^{(i)}_q$ and apply the BC procedure;
         \State broadcast $\{j: \mathsf{H}_j\ \text{rejected}\}$;
    \end{algorithmic}
    \label{alg:average BC}
\end{algorithm}
\begin{prop}
    Applying BC procedure to $(\overline{W}_j)_{j\in [m']}$ controls the FDR.
\end{prop}
\begin{IEEEproof}
    We must show that the conditional symmetry condition holds for $\left\{\overline{W}_j:\mathsf{H}_j\ \text{true}\right\}$. Let $\Hc_0=\left\{j:\mathsf{H}_j\ \text{true}\right\}$ We note
    %\footnote{In this equation with a slight abuse of notation we override the conditional probability measure $\P$ by the regular conditional probability measure $\P ^*$.}
    \begin{align*}
        \P &\left(\overline{W}_j > t \big |\,  \overline{W}_{-j} \right) \\ &=\E\left\{\P\left(\overline{W}_j > t \Big |\, W_{-j}^{(1)},\hdots, W_{-j}^{(N)}\right) \Big|\, \overline{W}_{-j}\right\}\\ 
        &=\E\left\{\P\left(-\overline{W}_j > t \Big |\, W_{-j}^{(1)},\hdots, W_{-j}^{(N)}\right) \Big|\, \overline{W}_{-j}\right\}\\
        &=\P \left(\overline{W}_j < -t \big |\,  \overline{W}_{-j} \right),\quad \forall j\in \Hc_0 \text{ and } \forall t\in\mathbb R .
    \end{align*}
This implies that 
\begin{equation*}
    \P \left(\overline{W}_j > 0 \Big |\, \left|\overline{W}_j\right |,  \overline{W}_{-j} \right)=\P \left(\overline{W}_j < 0 \Big |\, \left|\overline{W}_j\right |,  \overline{W}_{-j} \right)\ a.s.,
\end{equation*}
%$\P \left(\overline{W}_j > 0 \Big |\, \left|\overline{W}_j\right |,  \overline{W}_{-j} \right)=\P \left(\overline{W}_j > 0 \Big |\, \left|\overline{W}_j\right |,  \overline{W}_{-j} \right) a.s.$, 
completing the proof.
\end{IEEEproof}

\begin{figure*}[b]\centering
\begin{subfigure}{}\includegraphics[scale=0.44]{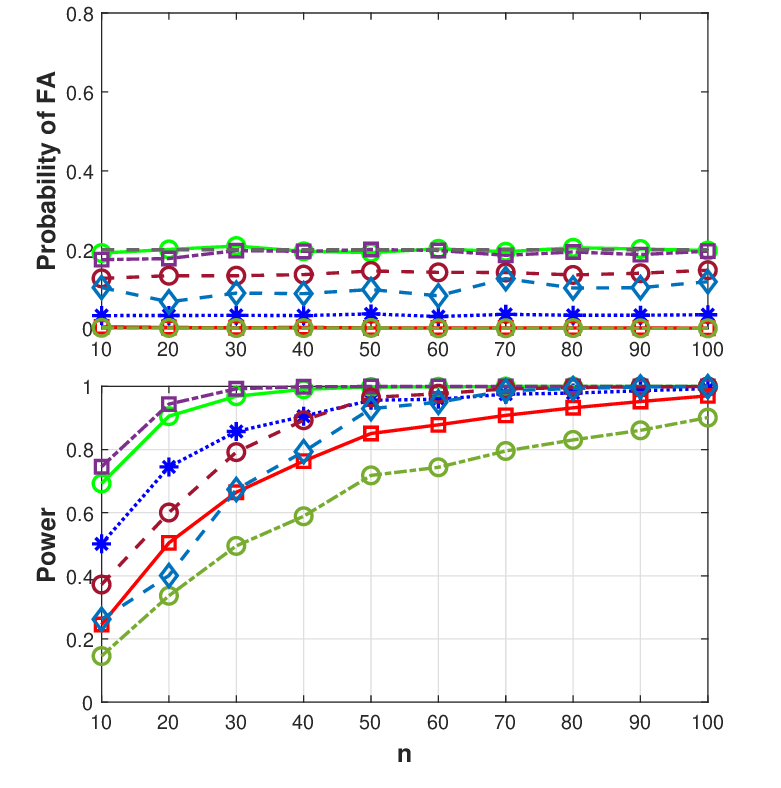}\end{subfigure}
\begin{subfigure}{}\includegraphics[scale=0.44]{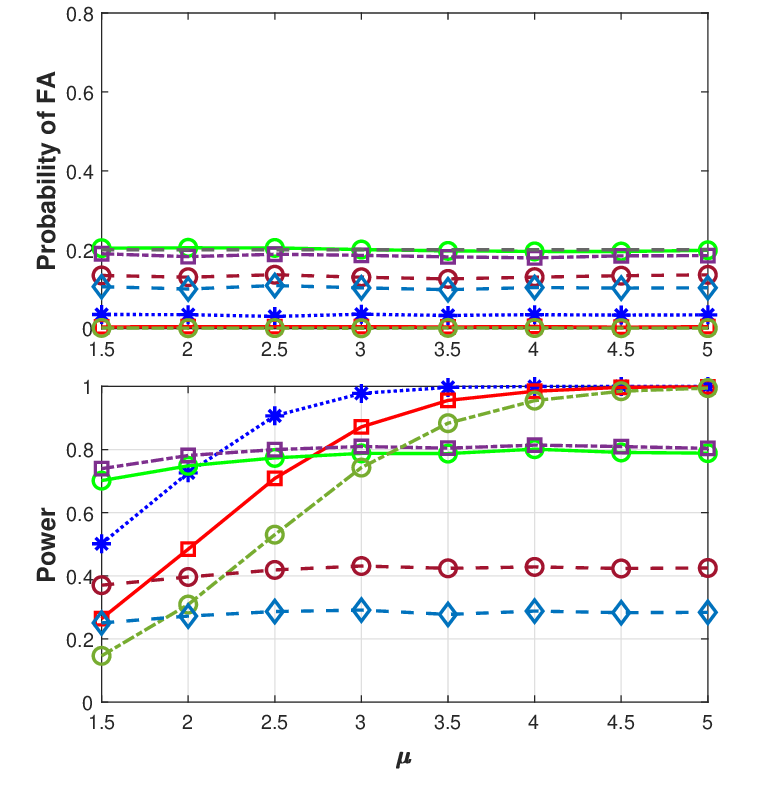}\end{subfigure}
\begin{subfigure}{}\includegraphics[height=1.9in]{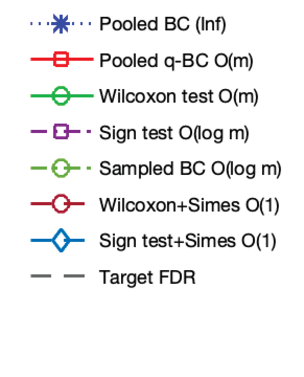}\end{subfigure}
\caption{Experiment~2. From left to right, Simulations I and II.}
\vspace{-2em}
\end{figure*}

\begin{figure*}[b]\centering
\begin{subfigure}{}\includegraphics[scale=0.4]{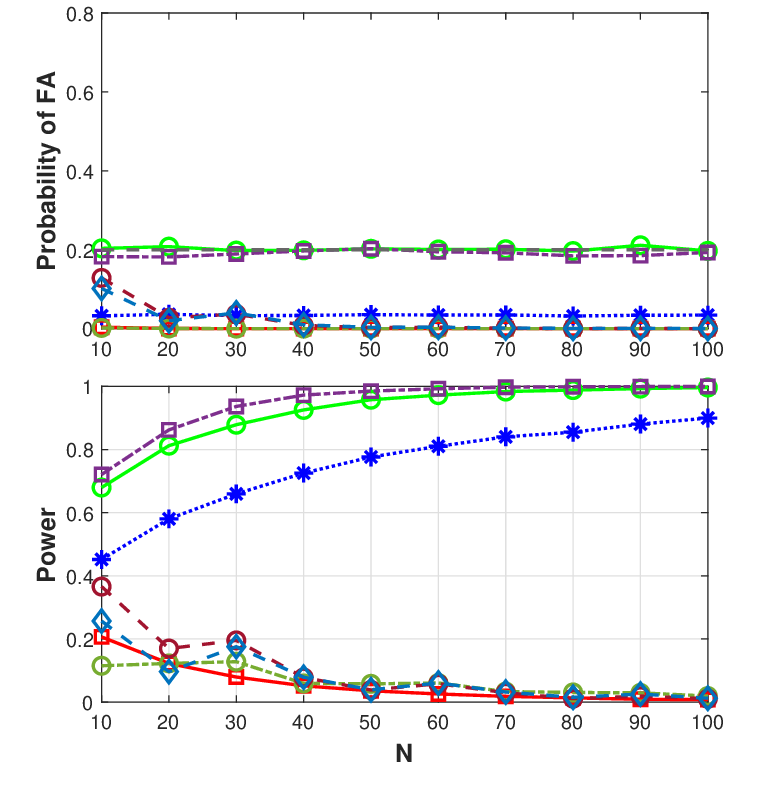}\end{subfigure}
\begin{subfigure}{}\includegraphics[scale=0.4]{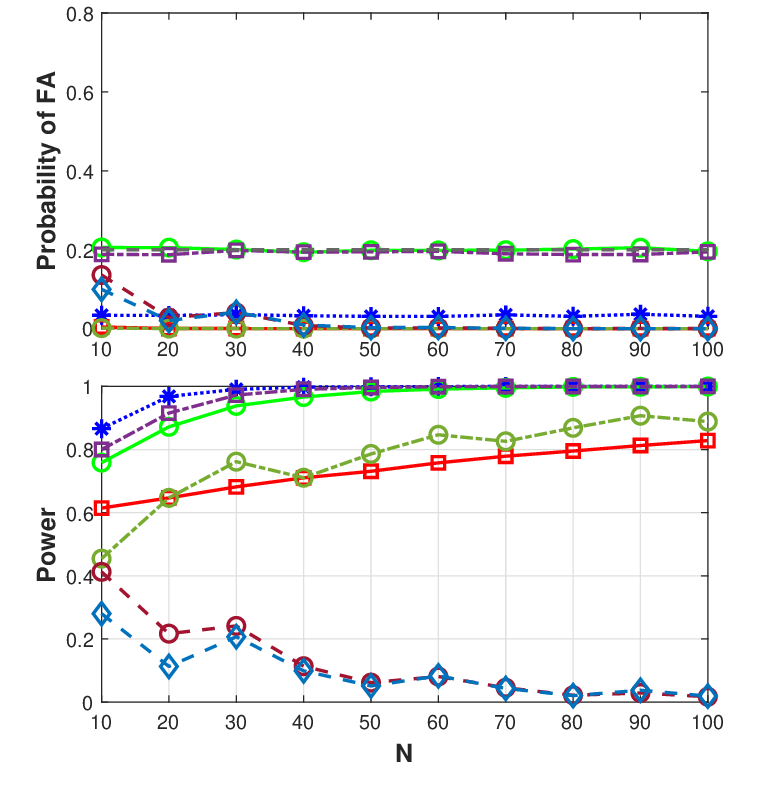}\end{subfigure}
\begin{subfigure}{}\includegraphics[scale=0.4]{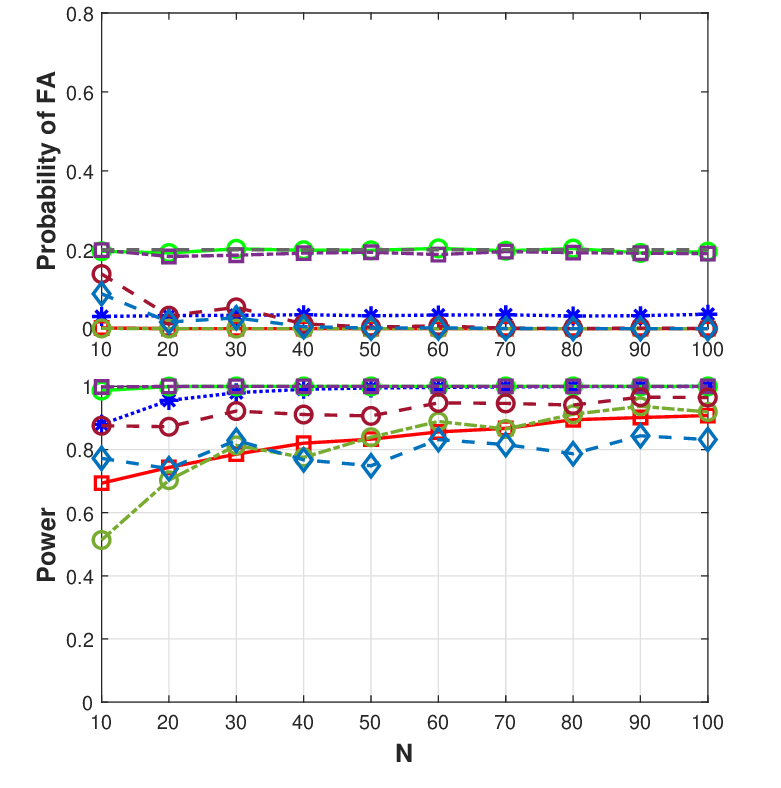}\end{subfigure}
\caption{Experiment~2. Simulation III: From left to right, $(n=10,\ \muv=1.5)$, $(n=10,\ \muv=2.5)$, and $(n=40,\ \muv=1.5)$.}
\vspace{-2em}
\end{figure*}

\begin{figure*}[t]\centering
\begin{subfigure}{}\includegraphics[scale=0.4]{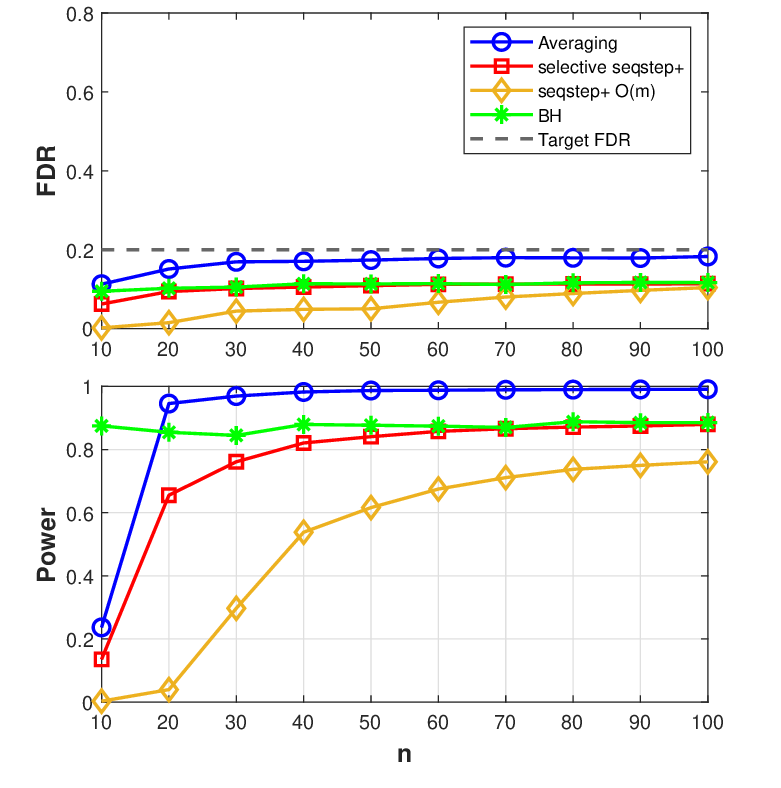}\end{subfigure}
\begin{subfigure}{}\includegraphics[scale=0.4]{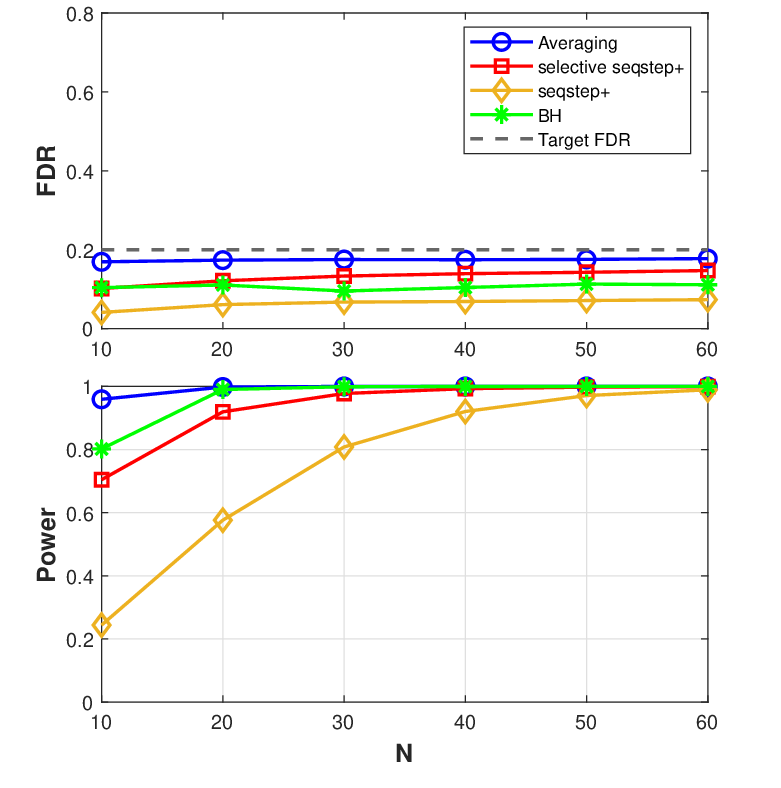}\end{subfigure}
\begin{subfigure}{}\includegraphics[scale=0.4]{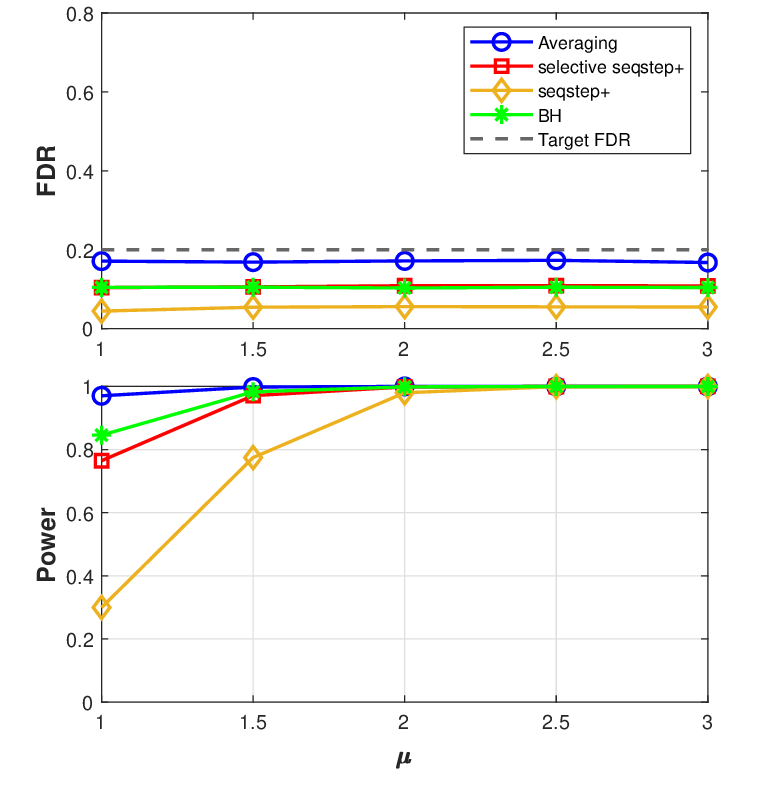}\end{subfigure}
\caption{ Experiment 3. From left to right, Simulations I, II, and III.}
\vspace{-2em}
\end{figure*}

\section{Simulation Study}
In this section, we evaluate and compare the empirical performance of our algorithms. In all the experiments, we have $n$ statistics at each node and $\alpha=0.2$. The number of false null hypotheses at node $i\in [N]$ is $\lfloor \pi_1^{(i)} n \rfloor$ where $\pi_1^{(i)}=0.3-0.2(i-1)/N$. At node $i$ the statistics are generated independently according to $\Norm(\mu,\sigma^2)$ where $\sigma^2 \sim \mathsf{Unif}[{\sigma^2}^{(i)}_{\text{base}}-0.25,{\sigma^2}^{(i)}_{\text{base}}+0.25]$, $\mu=0$ under null hypotheses and $\mu \sim \mathsf{Unif}[\mu_{\text{base}}^{(i)}-0.5,\mu_{\text{base}}^{(i)}+0.5]$ for false nulls. In all experiments, we set ${\sigma^2}^{(i)}_{\text{base}}=1+i/N$ for node $i\in[N]$. The FDR and power are estimated by averaging over $10000$ trials.

\smallskip
\noindent{{\bf Experiment 1 (individual decision).} In this experiment we wish to compare the two algorithms we have proposed for the individual decision problem, namely, pooled q-BC and sampled BC. The pooled BC procedure (i.e., the centralized BC procedure without any normalization, quanitization or sampling operations) is also presented as a baseline with an unlimited communication budget. We consider three simulation settings, namely, FDR and power vs number of statistics at each node $n$, number of nodes in the network $N$, and $\muv$ which determines $\mu_{\text{base}}^{(i)}$ via $\mu_{\text{base}}^{(i)}=\muv +i/N$.

\smallskip
{\noindent \underline{\textbf{Simulation I (vary $n$)}}}
We fix $N=10$, $\mu_{\text{base}}^{(i)}=2.5+i/N$ (i.e., $\muv = 2.5$), and $q=4$ levels of quantization for magnitudes in q-BC procedure. For the comparison to be fair, the number of samples in sampled BC procedure is computed by 
\begin{equation*}
    L = \left\lfloor\frac{1}{2}\frac{n\left(\lceil\log_2(q)\rceil+1\right)}{\lceil\log_2(n)\rceil}\right\rfloor.
\end{equation*}
In this case, both methods have (almost) the same bits of communication budget. Indeed, this is a little in favor of the pooled q-BC procedure since we take the floor to compute $L$ but it can be observed in Figure \ref{fig:individual} the sampled BC procedure outperforms the pooled q-BC procedure when $n\geq 20$.

\smallskip
{\noindent \underline{\textbf{Simulation II (vary $N$)}}}
In this simulation, we fix $n=50$, $\mu_{\text{base}}^{(i)}=2.5+i/N$, and $q=4$ levels of quantization for magnitudes in q-BC procedure. The number of samples for sampled BC procedure is determined according to the same formula as Simulation I. This simulation reveals that the difference in power observed in Simulation I vanishes as we increase the number of nodes in the network. 

\smallskip
{\noindent \underline{\textbf{Simulation III (vary $\muv$)}}}
In this simulation, we fix $N=10$, $n=50$, and $q=4$ levels of quantization for magnitudes in q-BC procedure. However, we let $\mu_{\text{base}}^{(i)}=\muv +i/N$ and vary $\muv$. The number of samples for sampled BC procedure is determined according to the same formula as Simulation I. This simulation shows that the difference in power observed in Simulation I remains persistent for a large range of $\muv$.
%vanishes when $\muv$ is large enough. 
 }

 \smallskip
\noindent{{\bf Experiment 2 (global decision)} In this experiment we compare the six algorithms we have proposed for the global decision problem. We fix $q=16$ and $L=5$, and consider five simulation settings, where FDR and power are plotted w.r.t. the number of statistics at each node $n$, $\muv$, and three are related to varying the number of nodes $N$ in the network.

\smallskip
{\noindent \underline{\textbf{Simulation I (vary $n$)}}}
In this simulation, we fix $N=10$, $\mu_{\text{base}}^{(i)}=1.5+i/N$, and vary the number of statistics at each node $n$. It can be observed that the sign test, even with an arguably low communication cost, outperforms the other methods. Also, the distributed Wilcoxon test (Wilcoxon+Simes) gives a considerably high power considering its extremely low communication cost. 

\smallskip
{\noindent \underline{\textbf{Simulation II (vary $\muv$)}}}
In this simulation, we fix $N=10$, $n=10$, and vary $\muv$ which determines $\mu_{\text{base}}^{(i)}$ via $\mu_{\text{base}}^{(i)}=\muv +i/N$. This simulation reveals that although the BC procedure (and its communication efficient variants) is designed to control FDR (which is a more complex type I error than the probability of false alarm), it still gives higher power than Wilcoxon test (the other $O(m)$ method) which has only asymptotic guarantee for the type I error.
 }

\smallskip
{\noindent \underline{\textbf{Simulation III (vary $N$)}}}
We consider three cases $(n=10,\ \muv=1.5)$, $(n=10,\ \muv=2.5)$, and $(n=40,\ \muv=1.5)$, and vary the number of nodes $N$ in the network. In the first case, we observe that increasing the number of nodes in the network results in power reduction for all methods except the sign and Wilcoxon tests. Regarding the second second case, where we allow relatively higher $\muv$, we observe that communication-efficient BC variants are powerful. Finally, in the third case, where we allow a relatively higher number of statistics at each node, all of the methods are powerful.

\smallskip
\noindent{{\bf Experiment 3 (distributed intersection hypothesis)} In this experiment we compare the algorithm we have proposed for the distributed intersection hypothesis problem with the method proposed in \cite{su2015communication} where intermediate p-values are computed through the communication of signs and the order is determined by quantized magnitudes. The paper only considers applying the Selective SeqStep in the center node. However, their method is compatible with BH procedure (which does not use any information from magnitudes; hence less communication cost) and the SeqStep procedure. We fix $q=16$, and consider three simulation settings, the same as Experiment~1. The observation is that the averaging method outperforms the other ones in all three simulations. 

\smallskip
{\noindent \underline{\textbf{Simulation I (vary $n$)}}}
We fix $N=10$, $\mu_{\text{base}}^{(i)}=1+i/N$. 

\smallskip
{\noindent \underline{\textbf{Simulation II (vary $N$)}}}
We fix $n=30$, $\mu_{\text{base}}^{(i)}=1+i/N$. The number of samples for sampled BC procedure is determined according to the same formula as in Simulation~I.

\smallskip
{\noindent \underline{\textbf{Simulation III (vary $\muv$)}}}
We fix $N=10$ and $n=30$. 

\section{Conclusion}
In many practical hypothesis testing settings, assuming the availability of the p-values might be restrictive in that it requires the knowledge of the distribution of the test statistics under the null. This work concerns distribution-free methods that only require a weak structural assumption (i.e., conditional symmetry) on the test statistics. We design, analyze, and evaluate communication-efficient procedures in three broad scenarios when such test statistics are available across a star network setting, following the appropriate type-I error metrics such as the FDR and global type-I error rate. 

 \balance
 %\clearpage
\bibliographystyle{IEEEtran}
\bibliography{ref.bib}

\end{document}